\documentclass[twocolumn,showpacs,pre,english,showkeys,aps,unsortedaddress,superscriptaddress,nofootinbib]{revtex4-2}
\usepackage{amsmath}
\usepackage[table, svgnames, dvipsnames]{xcolor}
\usepackage{siunitx}
\usepackage{amssymb}
\usepackage{epsfig}
\usepackage{graphicx}
\usepackage[colorlinks]{hyperref}
\usepackage[T1]{fontenc}
\usepackage[utf8]{inputenc}
\usepackage{float}
\usepackage{xcolor}
\usepackage[normalem]{ulem}
\usepackage{tikz}
\usetikzlibrary{positioning}

\begin{document}

\title{Crossover and universality breaking in the dilute Baxter–Wu model}

\author{Dimitrios Mataragkas}
\affiliation{School of Mathematics, Statistics and Actuarial Science, University of Essex, Colchester CO4 3SQ, United Kingdom}

\author{Alexandros Vasilopoulos}
\email{alex.vasilopoulos@essex.ac.uk}
\affiliation{School of Mathematics, Statistics and Actuarial Science, University of Essex, Colchester CO4 3SQ, United Kingdom}

\author{Dong-Hee Kim}
\email{dongheekim@gist.ac.kr}
\affiliation{Department of Physics and Photon Science, Gwangju Institute of Science and Technology, Gwangju 61005, Republic of Korea}

\author{Nikolaos G. Fytas}
\email{nikolaos.fytas@essex.ac.uk}
\affiliation{School of Mathematics, Statistics and Actuarial Science, University of Essex, Colchester CO4 3SQ, United Kingdom}

\date{\today}

\begin{abstract}
The critical behavior of the Baxter-Wu model belongs to the universality class of the four-state Potts model. While the introduction of annealed vacancies does not alter the criticality of the four-state Potts model, the dilute Baxter-Wu model has remained the subject of several competing scenarios. Here we investigate the phase diagram of the spin-$1$ Baxter-Wu model in the presence of a crystal field using transfer-matrix calculations and large-scale Monte Carlo simulations. Our results reveal a systematic evolution of the effective critical behavior with increasing crystal field, accompanied by increasingly strong finite-size corrections near the crossover to the first-order regime. Along the line of continuous transitions, the central charge remains close to $c=1$, while the scaling dimensions systematically deviate from the spin-$1/2$ limit as the crystal field increases, consistent with either continuously varying effective critical exponents or a slow crossover between competing critical behaviors. The first-order regime is independently characterized through multicanonical simulations, which confirm the expected finite-size scaling and interfacial behavior. Taken together, our results provide a unified picture of the dilute spin-$1$ Baxter-Wu model, substantially narrowing the range of possible scenarios for the crossover between continuous and first-order phase transitions.
\end{abstract}

\maketitle

\section{Introduction}
\label{sec:intro}

\begin{figure}[h]
    \centering
    \includegraphics[width=0.8\linewidth]{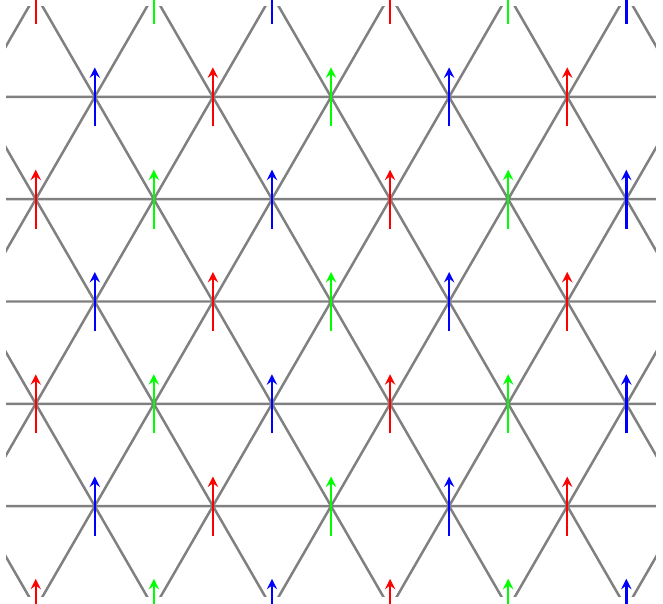}
    \caption{Representation of the Baxter-Wu triangular lattice as a superposition of the three sublattices, $A$, $B$, and $C$. Each sublattice corresponds to spins of a different color. The spins are shown in the ferromagnetic ground state.}
    \label{fig:lattice}
\end{figure}

The Baxter-Wu model occupies a special place in statistical physics as a rare example of a two-dimensional spin system with multispin interactions and broken spin-inversion symmetry~\cite{wood72} that is nevertheless exactly solvable~\cite{baxter73,baxter_book}. The pure spin-$1/2$ Baxter-Wu model is defined on the triangular lattice by the Hamiltonian
\begin{equation}\label{eq:Ham}
 \mathcal{H}^{\rm (pure)}
  = -J\sum_{\langle xyz \rangle}\sigma_{x}\sigma_{y}\sigma_{z},
\end{equation}
where the exchange interaction $J>0$, the sum runs over all elementary triangles of a lattice with $N$ sites, and $\sigma_x=\pm1$ are Ising spin-$1/2$ variables. The triangular lattice can be partitioned into three sublattices, $A$, $B$, and $C$, as illustrated in Fig.~\ref{fig:lattice}, such that each triangular face contains exactly one site from each sublattice. The ground state of the model is four-fold degenerate: one ferromagnetic configuration with all spins aligned and three ferrimagnetic configurations in which two sublattices carry spins pointing down and the remaining one spins pointing up. The Hamiltonian in Eq.~(\ref{eq:Ham}) is also self-dual~\cite{wood72,merlini72}, yielding the same critical temperature as the spin-$1/2$ Ising model on the square lattice, $k_{\rm B}T_{\rm c}/J = 2 / \ln{(1+\sqrt{2})} = 2.269185\ldots$,
where $k_{\rm B}$ denotes the Boltzmann constant. Its critical behavior is described by a conformal field theory with central charge $c=1$ and belongs to the universality class of the four-state Potts model~\cite{domany78}, albeit without the logarithmic corrections present in the latter~\cite{alcaraz97,alcaraz99}.  

A natural generalization introduces a crystal field (or single-ion anisotropy) coupling $\Delta$ and allows for a nonmagnetic state, leading to the spin-$1$ (or dilute) Baxter-Wu model. The Hamiltonian reads
\begin{equation}
\mathcal{H}^{\rm (dilute)} = -J \sum_{\langle xyz\rangle} \sigma_x \sigma_y \sigma_z + \Delta \sum_x \sigma_x^2 = E_{J}+\Delta E_{\Delta},
\end{equation}
where $\sigma_i=\{-1,0,1\}$, the sum extends over all elementary triangles of the triangular lattice, and $E_J$ and $E_\Delta$ the contributions of the exchange and the crystal field, respectively, to the total energy. Despite its apparent simplicity, this model is not exactly solvable and displays a rich phase diagram in the crystal-field--temperature $(\Delta,T)$ plane, including both continuous and first-order transition regimes, that have not been fully clarified~\cite{dias17,jorge21}. A sketch of the phase diagram is presented in the main panel of Fig.~\ref{fig:phase_diagram}; see also Table~\ref{table}.  

\begin{table*}[t]
\caption{Transition points and critical exponents of the spin-$1$ Baxter-Wu model obtained in the present work. The transition points from the transfer-matrix analysis are determined by extrapolation assuming first-order (TM1) and second-order (TM2) scaling forms, while the corresponding multicanonical (MUCA) estimates are listed in the third column. The last four columns compare the transfer-matrix estimates of the critical exponents $\eta$ and $\nu$ obtained using the two finite-size correction Ansätze discussed in the text.}
    \begin{ruledtabular}
    \begin{tabular}{S[table-format=2.4] S[table-format=1.6] S[table-format=1.8] S[table-format=3.1] S[table-format=1.6] S[table-format=1.8] S[table-format=1.8] S[table-format=1.8] S[table-format=1.8]}
         {$T$} & {$\Delta^*_\infty$ (TM1)} & {$\Delta^*_\infty$/$\Delta_c$ (MUCA)} & {$\Delta$} & {$T_{\rm c}$ (TM2)} & {$\eta^\mathrm{eff}_\infty$ [Eq.~\eqref{eq:RM12}]} & {$\eta^\mathrm{eff}_\infty$ [Eq.~\eqref{eq:RM24}]} & {$\nu^\mathrm{eff}_\infty$ [Eq.~\eqref{eq:RM12}]} & {$\nu^\mathrm{eff}$ [Eq.~\eqref{eq:RM24}]} \\
         \hline
         0.30 & 1.999236 & & & & & & & \\
         0.35 & 1.997686 & & & & & & & \\
         0.40 & 1.994580 & & & & & & & \\
         0.45 & 1.989318 & & & & & & & \\
         0.50 & 1.981347 & & & & & & & \\
         0.55 & 1.970175 & & & & & & & \\
         0.60 & 1.955361 & & & & & & & \\
         0.65 & 1.936497 & & & & & & & \\
         0.70 & 1.913193 & & & & & & & \\
         0.75 & 1.885062 & & & & & & & \\
         0.80 & 1.851714 & & & & & & & \\
         0.85 & 1.812755 & & & & & & & \\
         0.90 & 1.767781 & & & & & & & \\
         0.95 & 1.716384 & & & & & & & \\
         1.00 & 1.658160 & & & & & & & \\
         1.05 & 1.592709 & & & & & & & \\
         1.10 & 1.519634 & 1.51966(2) & & & & & & \\
         1.15 & 1.438523 & & & & & & & \\
         1.20 & 1.348935 & & & & & & & \\
         1.25 & 1.250375 & & & & & & & \\
         1.30 & 1.142281 & & & & & & & \\
              &          & &   1.0 & 1.359662 & 0.2055(4)  & 0.2103(3)  & 0.0591(3)  & 0.597(2) \\
              &          & &   0.9 & 1.398081 & 0.2150(3)  & 0.2182(2)  & 0.602(2)   & 0.607(1) \\
         1.40 & 0.894766 & & & & & \\
              &          & &   0.8 & 1.434077 & 0.2217(2)  & 0.2239(1)  & 0.611(1)   & 0.615(1) \\
              &          & &   0.7 & 1.467937 & 0.2267(1)  & 0.2283(1)  & 0.618(1)   & 0.6215(8) \\
              &          & &   0.6 & 1.499896 & 0.23053(6) & 0.23170(6) & 0.6249(8)  & 0.6272(6) \\
         1.50 & 0.599658 & & & & & & &\\
         1.5301 & & 0.4999(2) & & & & & & \\
              &          & &   0.5 & 1.530149 & 0.23350(3) & 0.23437(5) & 0.6301(6)  & 0.6319(4) \\
              &          & &   0.4 & 1.558856 & 0.23586(2) & 0.23653(3) & 0.6345(5)  & 0.6360(3) \\
              &          & &   0.3 & 1.586152 & 0.23777(1) & 0.23830(2) & 0.6383(4)  & 0.6394(3) \\
              &          & &   0.2 & 1.612155 & 0.23934(1) & 0.23976(1) & 0.6416(3)  & 0.6424(2) \\
              &          & &   0.1 & 1.636964 & 0.24065(1) & 0.24099(1) & 0.6443(2)  & 0.6450(2) \\
         1.6606 & & 0.0008(7) & & & & & & \\
              &          & &   0.0 & 1.660667 & 0.24175(1) & 0.24204(1) & 0.6468(2)  & 0.6473(1) \\
              &          & &  -1.0 & 1.850262 & 0.24711(1) & 0.24723(1) & 0.65955(7) & 0.65954(2)  \\
              &          & & -10.0 & 2.257751 & 0.24986(1) & 0.24997(1) & 0.66678(5) & 0.666664(1) \\
    \end{tabular}
    \end{ruledtabular}
    \label{table}
\end{table*}

Based on analogies with diluted Potts models~\cite{nienhuis79} and supporting numerical evidence, the existence of a multicritical point at finite values of $\Delta$ has long been conjectured~\cite{costa04}. However, both its location and even its nature remain unsettled~\cite{kinzel81,dias17,jorge21}, with additional conflicting scenarios regarding universality and other transition characteristics being reported in the literature~\cite{costa04b,costa16,jorge21}. In Ref.~\cite{dias17}, the location of a pentacritical point was estimated as $(\Delta_{\rm pp},T_{\rm pp})\approx (0.8902,1.4)$, whereas Ref.~\cite{jorge21} proposed the substantially different values $(\Delta_{\rm pp},T_{\rm pp})\approx [1.68288(62),0.98030(10)]$. If present, this pentacritical point corresponds to the coexistence of three ferrimagnetic configurations, a ferromagnetic configuration, and the zero-spin state. Furthermore, the validity of universality along the continuous transition line remains under debate~\cite{vasilopoulos22,macedo23,macedo24}, with proposals ranging from four-state Potts criticality to continuously varying critical exponents~\cite{costa04b,costa16,jorge21}. An outstanding question is whether the reported deviations from four-state Potts behavior reflect a genuine modification of the asymptotic critical behavior or originate from strong crossover effects and finite-size corrections in the vicinity of the first-order regime. With respect to the first-order transition regime, to the best of our knowledge no relevant study currently exists. 

We revisit this longstanding problem using complementary numerical and theoretical approaches. By combining transfer-matrix calculations, finite-size scaling, and large-scale Monte Carlo simulations, we systematically explore the phase transitions of the dilute spin-$1$ Baxter–Wu model across the $(\Delta,T)$ phase diagram. Our results reveal a line of continuous transitions extending towards a broad crossover region and provide no numerical evidence for an isolated multicritical point within the resolution of the present study. Along the continuous transition line, central-charge estimates remain close to $c\simeq1$, while both transfer-matrix and Monte Carlo analyses reveal a systematic evolution of the effective scaling dimensions away from the four-state Potts values~\cite{domany78}. As the crossover region is approached, finite-size effects become apparent, making the extraction of asymptotic critical behavior progressively more difficult. The observed evolution is compatible with either continuously varying effective critical exponents or a slow crossover between competing critical behaviors, and our results do not allow us to unambiguously distinguish between these two scenarios. In the intermediate region, previously conjectured to host a multicritical point, we instead identify an extended crossover regime characterized by pronounced finite-size effects and continuously evolving effective conformal properties. This behavior suggests that the crossover from continuous to first-order transitions occurs smoothly over a finite interval of crystal-field values rather than through an isolated higher-order singular point. Finally, in the low-temperature (large-$\Delta$) regime, we establish the first-order nature of the transition through independent evidence from multicanonical simulations, including phase coexistence, finite interfacial tension, and the expected volume scaling of thermodynamic response functions.

The remainder of this paper is organized as follows. In Sec.~\ref{sec:TM} we present an extensive transfer-matrix analysis of the dilute spin-$1$ Baxter-Wu model, focusing on the evolution of the conformal properties and effective scaling dimensions along the continuous transition line and across the proposed crossover region. Section~\ref{sec:MC} complements this study with large-scale Monte Carlo simulations, where finite-size scaling and multicanonical methods are employed to investigate the critical behavior, establish the first-order character of the low-temperature regime, and assess the crossover between the two regimes. Finally, in Sec.~\ref{sec:conclusions} we summarize our results, discuss their implications for the universality of the dilute Baxter-Wu model, and outline several open questions concerning the nature of the crossover and its possible renormalization-group interpretation.

\section{Transfer-matrix analysis}
\label{sec:TM}

In the limit $\Delta=-\infty$, the spin-$1$ model reduces to the spin-$1/2$ Baxter-Wu model. Within conformal field theory~\cite{alcaraz97}, the spectrum on an infinite strip of finite width $M$ with periodic boundary conditions is described by
\begin{equation} 
\label{eq:CFT}
\epsilon_\alpha(M) = \epsilon_\infty + \frac{2\pi v_s}{M^2} 
\left( x_\alpha - \frac{c}{12} + R_\alpha(M) \right),
\end{equation}
where $\epsilon_\alpha = -(\ln \lambda_\alpha)/M$ is the energy associated with the transfer-matrix eigenvalue $\lambda_\alpha$. 
The central charge $c$ and the scaling dimensions $x_\alpha$ can be estimated by analyzing the finite-size behavior of the spectrum. In particular, the effective central charge is obtained from the ground-state energies as
\begin{equation} \label{eq:c_eff}
c_{\mathrm{eff}}(M) = \frac{6}{\pi v_s}
\frac{\epsilon_0(M+3) - \epsilon_0(M)}{M^{-2} - (M+3)^{-2}},
\end{equation}
where the ground state corresponds to $x_0 = 0$. Similarly, the effective scaling dimensions are given as
\begin{equation} \label{eq:xeff}
x^\mathrm{eff}_\alpha(M) = \frac{1}{2\pi v_s} 
M \ln \frac{\lambda_0}{\lambda_\alpha},
\end{equation}
whose convergence is governed by the finite-size corrections $R_\alpha(M)$. It is well established that the spin-$1/2$ model belongs to the four-state Potts universality class~\cite{domany78}, characterized by a central charge $c=1$ and scaling dimensions $x_\alpha \in \{0,\,1/8,\,1/2,\ldots\}$. However, the finite-size corrections differ: for the spin-$1/2$ Baxter-Wu model they scale as $R_\alpha(M)\sim M^{-2}$, in contrast to the logarithmic corrections expected for the four-state Potts model~\cite{alcaraz97}.

\begin{figure}
   \centering
   \includegraphics[width=1.0\linewidth]{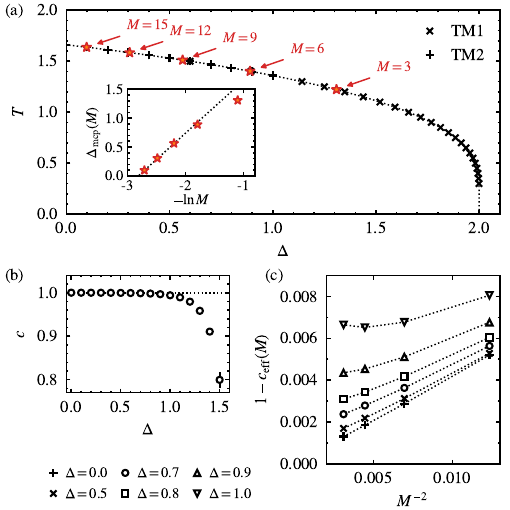}
   \caption{Phase diagram of the spin-$1$ Baxter-Wu model. 
   (a) Transition points in the $(\Delta,T)$ plane, shown for $\Delta \ge 0$. The labels TM1 and TM2 denote the transitions points located using the scaling form for the first-order and second-order transitions, respectively~\cite{SM}. Stars indicate the heuristic test for a multicritical point proposed in Ref.~\cite{dias17}, based on size triplets $(M,M+3,M+6)$. (b) Central charge $c$ estimated along the transition line. (c) Finite-size behavior of the effective central charge for $M \ge 9$. Dotted lines serve as guides to the eye.}
   \label{fig:phase_diagram}
\end{figure} 

A possible scenario for finite $\Delta$ is that the critical exponents vary continuously along the critical line until it meets the first-order regime at a multicritical point, as suggested by early transfer-matrix calculations for $M\leq9$~\cite{costa04}. An alternative possibility is that the observed evolution of the critical properties reflects a broad crossover accompanied by increasingly strong finite-size effects as the first-order regime is approached. We revisit these scenarios by extending the calculations up to $M=21$ within our available computational resources, employing sparse-matrix factorization in the geometry with $v_s=\sqrt{3}$~\cite{mataragkas25b}. 

In particular, we construct the transfer matrix $\mathbf{T}$ in the two-layer geometry of the triangular lattice following the sparse-matrix factorization introduced in Ref.~\cite{mataragkas25b}. We adopt the same convention for labeling sites in the strip geometry illustrated in Fig.~2 of Ref.~\cite{mataragkas25b}, except that here $M$ denotes the strip width, to avoid confusion with the linear system size $L$ used in the Monte Carlo simulations on $L\times L$ lattices. The transfer matrix can be written in the factorized form
\begin{equation}
    \mathbf{T} = \mathbf{R} 
    \left( \mathbf{V}^{1/2}
    \mathbf{T}_M \mathbf{T}_{M-1} \cdots \mathbf{T}_2 \mathbf{T}_1
    \mathbf{V}^{1/2} \right)^2 ,
\end{equation}
where $\mathbf{V}$ is a diagonal matrix with elements
\begin{equation}
    \langle \mathbf{s} | \mathbf{V} | \mathbf{s}' \rangle =
    \exp\!\left[ -\beta \Delta \sum_{j=1}^{M} s_j^2 \right]
    \prod_{j=1}^{M} \delta_{s_j,s'_j}.
\end{equation}
The matrix elements of $\mathbf{T}_1$, $\mathbf{T}_{k}$ ($k=2,\ldots,M-1$), and $\mathbf{T}_M$ are given by
\begin{align}
\langle \mathbf{s} | \mathbf{T}_1 | \mathbf{s}' \rangle
&=
\exp\!\left(\beta J s_1 s'_1 s'_2\right)
\delta_{s_{M+1},s'_1}
\prod_{j=2}^{M}\delta_{s_j,s'_j},
\\
\langle \mathbf{s} | \mathbf{T}_k | \mathbf{s}' \rangle
&=
\exp\!\left[
\beta J s_k s'_k
\left(s_{k-1}+s'_{k+1}\right)
\right]
\nonumber\\
&\qquad\times
\prod_{j=1}^{k-1}\delta_{s_j,s'_j}
\prod_{j=k+1}^{M+1}\delta_{s_j,s'_j},
\\
\langle \mathbf{s} | \mathbf{T}_M | \mathbf{s}' \rangle
&=
\exp\!\left[
\beta J s_M
\left(
s_{M-1}s'_M
+s'_Ms'_{M+1}
+s_1s'_{M+1}
\right)
\right]
\nonumber\\
&\qquad\times
\prod_{j=1}^{M-1}\delta_{s_j,s'_j},
\end{align}
respectively. In the final step, $\mathbf{R}$ denotes a cyclic rotation operator acting on the site indices, which restores the symmetry of the transfer matrix. To obtain the leading part of the spectrum, we employ the thick-restart Lanczos algorithm and compute the five largest eigenvalues $(\lambda_{0,1,2,3,4})$. These eigenvalues allow us to track the effective central charge $c_\mathrm{eff}$ and the effective scaling dimensions $x^\mathrm{eff}_1$ and $x^\mathrm{eff}_4$. Here $\lambda_{1,2,3}$ are degenerate by symmetry, which requires $M$ to be a multiple of $3$. 

Transition points are determined using two complementary finite-size procedures. Along the continuous transition line, we locate the crossing temperature $T_{M,M+3}$ of the scaled correlation length
$\xi_M/M=[M\ln(\lambda_0/\lambda_1)]^{-1} \propto 1/x_1^\mathrm{eff}$ for strips of widths $M$ and $M+3$. Assuming the finite-size expansions
\begin{equation}
\xi_M(T_{\rm c}) = A_0M\left(1+A_1M^{-1}+A_2M^{-2}+\cdots\right),
\end{equation}
and
\begin{equation}
\left(\frac{{\rm d}\xi_M}{{\rm d}T}\right)_{T_{\rm c}} = 
B_0M^{1+y}\left(1+B_1M^{-1}+B_2M^{-2}+\cdots\right),
\end{equation}
the crossing temperatures are extrapolated according to
\begin{equation}
T_{M,M+3}=T_{\rm c}+M^{-y}\left(aM^{-1}+bM^{-2}\right),
\end{equation}
where $y=1/\nu+\tilde{\omega}$ is treated as an effective correction exponent~\cite{dias17,jung17}. For $\Delta = \{-10,-1,0,0.5 \}$, this procedure yields $T_{\rm c} = \{2.257751,1.850262,1.660667,1.530149\}$, in excellent agreement with the Monte Carlo estimates reported in Ref.~\cite{macedo23}.
Within the first-order regime, transition points are instead determined from the minimum of the spectral gap between $\epsilon_0$ and $\epsilon_4$, corresponding to the effective scaling dimension $x_4^{\rm eff}(M)$. Near an avoided spectral crossing, the resulting sequence $\Delta_M$ converges exponentially with increasing strip width~\cite{jung17,mataragkas25b}, whereas closer to the crossover region stronger finite-size effects require an empirical power-law extrapolation. Despite relying on different assumptions regarding the order of the transition, the two approaches yield mutually consistent estimates throughout the crossover region shown in Fig.~\ref{fig:phase_diagram}(a), despite the absence of unambiguous multicritical signatures.

In estimating the central charge, we find that extrapolations of $c_\mathrm{eff}$ using data for $M \ge 9$ yield values very close to $c=1$ for $\Delta \lesssim 0.5$. For example, at $\Delta=0.5$ we obtain $c=0.9997(1)$, after which the estimates gradually decrease, as shown in Fig.~\ref{fig:phase_diagram}(b). Identifying the point at which the conformal description ceases to apply remains challenging, as the finite-size corrections become increasingly pronounced upon approaching the crossover region. However, the finite-size behavior of $c_\mathrm{eff}$ in Fig.~\ref{fig:phase_diagram}(c) reveals a systematic deviation from the expected correction $R_0\sim M^{-2}$ as $\Delta$ increases. In particular, the pronounced curvature observed at $\Delta=1$ indicates that the standard conformal field theory prediction %scaling Ansatz 
no longer provides a satisfactory description of the data, suggesting that the system is leaving the regime governed by the %$c \simeq 1$ 
critical theory. This observation is fully consistent with the independent Monte Carlo evidence presented below for the emergence of first-order behavior.  

\begin{figure}
    \centering
    \includegraphics[width=1.0\linewidth]{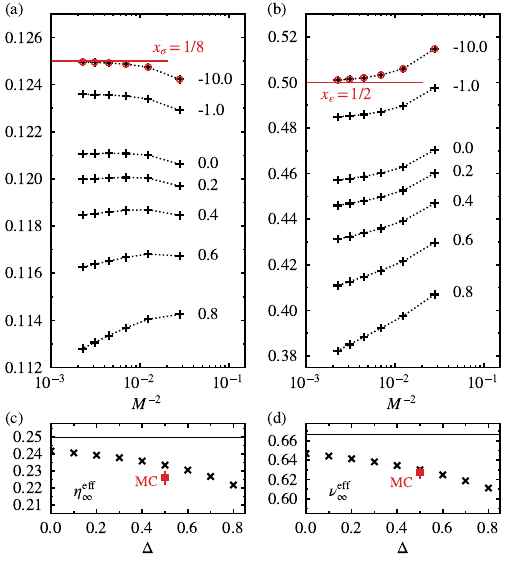}
    \caption{Scaling dimensions with increasing dilution. Finite-size behavior of the effective scaling dimensions $x^\mathrm{eff}_1$ (a) and $x^\mathrm{eff}_4$ (b) for several values of $\Delta$ indicated in the panels. Open circles denote results for the spin-$1/2$ model. The corresponding effective critical exponents (c) $\eta^\mathrm{eff}$ and (d) $\nu^\mathrm{eff}$ extrapolated to $M=\infty$ are compared with Monte Carlo (MC) estimates at $\Delta=0.5$ (red squares). Solid lines indicate the exact values of the four-state Potts universality class, $\eta=1/4$ and $\nu=2/3$.}
    \label{fig:CFT}
\end{figure}

On the other hand, establishing $c=1$  
is not sufficient to identify the universality class of the transition, since distinct critical theories may share the same central charge while differing in their operator content. In particular, there is no clear evidence that the scaling dimensions recover those of the spin-$1/2$ model at finite $\Delta$. Figure~\ref{fig:CFT} shows that the effective scaling dimensions $x^{\mathrm{eff}}_1$ and $x^{\mathrm{eff}}_4$ drift systematically away from the spin-$1/2$ values $x_\sigma=1/8$ and $x_\epsilon=1/2$ as $\Delta$ increases. Notably, for a fixed $\Delta$, the estimate of $x^{\mathrm{eff}}_4$ moves even farther from the spin-$1/2$ line as the strip width $M$ increases, indicating that the observed deviation cannot be attributed solely to the finite strip widths accessible in the transfer-matrix calculations. At the same time, the increasingly pronounced finite-size corrections observed as the crossover region is approached prevent a definitive determination of the asymptotic scaling behavior. Nevertheless, the systematic evolution of both scaling dimensions with increasing $\Delta$, together with their excellent agreement with the corresponding Monte Carlo estimates presented below, provides compelling evidence that the effective critical behavior evolves continuously along the transition line. Whether this evolution reflects a genuine line of critical fixed points or a slow renormalization-group crossover between competing critical behaviors remains an open question beyond the resolution of the present study.

The coexistence of a central charge close to $c=1$ with scaling dimensions that unfold away from the four-state Potts values is reminiscent of several exactly solved two-dimensional models possessing critical lines, most notably the Ashkin-Teller and eight-vertex models~\cite{ashkin43,baxter_book}. Along those critical lines, the central charge remains fixed at $c=1$, while the scaling dimensions vary continuously as a function of a marginal coupling. From this perspective, the behavior observed in the dilute Baxter-Wu model may suggest a similar mechanism. However, an important distinction is that in the Ashkin-Teller and eight-vertex models the continuously varying exponents are understood within a well-established conformal-field-theory framework, whereas no corresponding field-theoretical description is presently available for the dilute Baxter-Wu model. Moreover, the increasingly pronounced finite-size effects observed as the crossover regime is approached leave open the possibility that the apparent variation of the exponents reflects a slow renormalization-group crossover rather than a genuine line of fixed points. Establishing the precise theoretical relation between these scenarios remains an interesting open problem.

To estimate the effective critical exponents, $\eta^\mathrm{eff} \equiv 2x^{\mathrm{eff}}_1$ and $\nu^\mathrm{eff} \equiv (2-x^{\mathrm{eff}}_4)^{-1}$, we extrapolate the effective scaling dimensions to the limit $M \to \infty$ by assuming the finite-size correction
\begin{equation} \label{eq:RM12}
R_\alpha(M) \simeq a_\alpha M^{-1} + b_\alpha M^{-2},
\end{equation}
where the $M^{-2}$ term is inherited from the spin-$1/2$ limit. Table~\ref{table} lists the estimates obtained using data for $M \ge 9$. The additional $M^{-1}$ contribution is introduced to capture the pronounced curvature of $x^\mathrm{eff}_1$ for $\Delta \gtrsim 0$, which may also explain the inconclusive search for a multicritical point based on a plateau of $x^\mathrm{eff}_1$ as a heuristic indicator~\cite{dias17}. As illustrated in Fig.~\ref{fig:CFT}(a), increasing $M$ in the size-triplet test systematically shifts the apparent plateau toward smaller $\Delta$. This behavior can be interpreted in terms of the drift of the zero-slope point $M=-2b/a$ of $R_\alpha(M)$. The ratio $|a/b|$ decreases as $\Delta$ becomes more negative, recovering the expected scaling $R_\alpha(M)\propto M^{-2}$ in the limit $\Delta = -\infty$. For example, at $\Delta=-10$ we obtain $|a_1/b_1|<10$ and $|a_4/b_4|<50$, yielding $\eta^{\mathrm{eff}}_\infty=0.24986(1)$ and $\nu^{\mathrm{eff}}_\infty=0.66678(5)$, in excellent agreement with the exact $1/8$ and $2/3$ exponents of the spin-$1/2$ limit. The gradual evolution of the exponents is quantified by the extrapolations shown for $\Delta > 0$ in Figs.~\ref{fig:CFT}(c) and (d).

The tendency toward continuously varying effective critical exponents with increasing $\Delta$ remains robust under alternative choices of the finite-size correction Ansatz. In particular, we also considered the conventional correction form expected for the spin-$1/2$ Baxter-Wu model~\cite{alcaraz97},
\begin{equation}
\label{eq:RM24}
R_\alpha(M) \simeq a_\alpha M^{-2} + b_\alpha M^{-4}.
\end{equation}
The additional $M^{-4}$ term accounts for the weak curvature of the finite-size data and provides satisfactory fits for $M \ge 12$. As summarized in Table~\ref{table}, this alternative extrapolation leads to the same qualitative conclusion: the effective critical exponents evolve systematically away from their spin-$1/2$ values as $\Delta$ increases. We therefore conclude that the observed drift of the exponents is not an artifact of the particular extrapolation scheme employed. At the same time, we emphasize that both extrapolation procedures remain empirical, and establishing the asymptotic finite-size correction structure of the spin-$1$ Baxter-Wu model will require further theoretical developments together with transfer-matrix calculations for substantially larger strip widths.

A remaining question concerns the nature of the crossover from continuous to first-order behavior and, in particular, whether it is mediated by an isolated multicritical point or extends over a finite region of the phase diagram. While the departure from $c = 1$ observed in Fig.~\ref{fig:phase_diagram}(c) suggests that the onset of this crossover occurs below $\Delta = 1$, the resolution of the transfer-matrix calculations is ultimately limited by the accessible strip widths $M$ and the increasingly strong finite-size corrections observed in this regime. It is instructive to contrast this behavior with the Blume-Capel model, where the tricritical point is characterized by a distinct set of central charge and scaling dimensions, enabling precise finite-size scaling analyses via the persistence length~\cite{mataragkas25b}. In addition, the finite-size correction of $x^\mathrm{eff}_1$ in the Blume-Capel model exhibits a sharp sign change across the first- and second-order regimes, making the heuristic plateau test particularly effective. By contrast, the transfer-matrix spectrum of the spin-$1$ Baxter-Wu model develops smoothly, without exhibiting comparably robust spectral signatures that could be uniquely associated with a multicritical fixed point. Rather than providing direct evidence for an isolated multicritical point, the transfer-matrix results indicate a gradual evolution of the effective conformal properties accompanied by increasingly pronounced finite-size effects. From a renormalization-group perspective, such behavior is not expected for a conventional isolated multicritical fixed point, whose influence should become confined to an increasingly narrow critical region under coarse graining. Instead, it is naturally interpreted as evidence for an extended crossover preceding the onset of first-order behavior. Distinguishing between these possibilities requires complementary Monte Carlo evidence probing substantially larger length scales than those accessible within the transfer-matrix approach.

\begin{figure}
    \centering
    \includegraphics[width=1.0\linewidth]{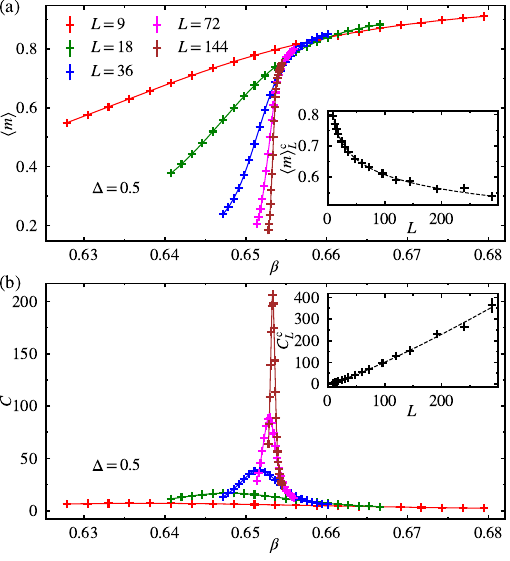} 
    \caption{(a) Order-parameter $\langle m \rangle$ and (b) specific-heat $C$ (b) curves as functions of the inverse temperature ($\beta \equiv 1/T$) obtained from Monte Carlo simulations. Representative system sizes are indicated in the panels. Insets show the corresponding finite-size scaling at the critical point for $\Delta=0.5$.}
    \label{fig:second-order}
\end{figure}

\section{Monte Carlo simulations}
\label{sec:MC}

To complement the transfer-matrix analysis within the continuous transition regime, we implement large-scale Metropolis Monte Carlo simulations combined with histogram reweighting~\cite{barkema_book}. We investigate the model at $\Delta = 0.5$ for linear system sizes $L \in \{9 - 288\}$. For each system size an average over $100$ independent realizations is performed to increase statistical accuracy. All runs are carried out in reduced units ($J=1$, $k_{\rm B}=1$) on triangular lattices with periodic boundary conditions. To accommodate both the ferromagnetic ground state and the three ferrimagnetic configurations, the linear system size $L$ is restricted to multiples of three~\cite{costa16}. The sampled observables include estimates of the mean energy $\langle E \rangle$, the order parameter $\langle m\rangle$ which is estimated from  the root mean square average of the magnetization per site of the three sublattices A, B, and C (see Fig.\ref{fig:lattice} and Refs.~\cite{costa04b,costa16,jorge20})
\begin{equation}\label{eq:order-parameter}
	m = \sqrt{\frac{m_{\rm A}^{2} +m_{\rm B}^{2} +m_{\rm C}^{2}}{3}},
\end{equation}
and the specific heat
\begin{equation}\label{eq:spec-heat}
	C = \left[\langle E^{2}\rangle - \langle E \rangle^{2}\right]/(LT)^{2}.
\end{equation}
Finally, for all scaling analyses, fits are performed using data with $L \ge L_{\rm min}$, and their quality was assessed using the standard $\chi^{2}$ goodness-of-fit test. In particular, fits are considered acceptable only when the quality-of-fit parameter satisfied $10\% < Q < 90\%$~\cite{press92}.

Typical order-parameter ($\langle m \rangle$) and specific-heat ($C$) curves for representative system sizes are shown in the main panels of Fig.~\ref{fig:second-order}. The insets display the corresponding finite-size scaling behavior at the critical point $\beta_{\rm c}=0.653562$ for $\Delta=0.5$~\cite{macedo23}, following the forms $\langle m \rangle_{L}^{\rm c} = b_{m} L^{-\beta/\nu}$ and $C_{L}^{\rm c} = b_{C} L^{\alpha/\nu}$, where $\{b_{m}, b_{C} \}$ are non-universal amplitudes. From the scaling of the order parameter [Fig.~\ref{fig:second-order}(a)] we obtain $\eta=0.226(4)$ via the Fisher scaling relation~\cite{landau_book}, $\eta = 2(\beta/\nu)-(d-2)$. The analysis of the specific heat [Fig.~\ref{fig:second-order}(b)] yields $\alpha/\nu=1.188(5)$, which in turn gives $\nu = 2/(d+\alpha/\nu) = 0.627(3)$ through the hyperscaling relation. The Monte Carlo estimates of $\eta$ and $\nu$ closely track the transfer-matrix effective exponents [see Figs.~\ref{fig:CFT}(c) and (d)] and differ from the four-state Potts values $\eta = 1/4$ and $\nu = 2/3$. These deviations become increasingly pronounced as the crossover region is approached and indicate a systematic evolution of the effective critical behavior along the transition line. At the same time, the growing importance of finite-size effects suggests that the extraction of the asymptotic critical behavior becomes progressively more challenging in this regime, and therefore a crossover interpretation cannot presently be excluded. This behavior contrasts with the Blume-Capel ferromagnet~\cite{zierenberg17,blume66}, where universality remains robust along the second-order transition line.

A natural question concerns the microscopic origin of the apparent departure from four-state Potts universality. Although the present results do not allow us to identify a unique mechanism, they suggest that the interplay between dilution and the three-spin interaction plays a central role. Unlike random dilution of pairwise interactions, removing a single spin simultaneously eliminates several elementary three-spin plaquettes, producing correlated local perturbations to the interaction network. As the crystal field increases, dilution locally suppresses elementary triangular interactions in a correlated manner, modifying the effective connectivity of the interaction network while preserving the fourfold degeneracy of the ordered phase. This mechanism differs qualitatively from annealed dilution in the four-state Potts model, where universality is known to remain unchanged~\cite{nienhuis79}. Our observations are therefore consistent with the picture that the combination of multispin interactions and dilution generates additional relevant scaling fields, leading to pronounced crossover effects and, possibly, continuously varying effective critical behavior before the system ultimately crosses over to a first-order transition.

\begin{figure}[ht!]
    \centering
    \includegraphics[width=1.0\linewidth]{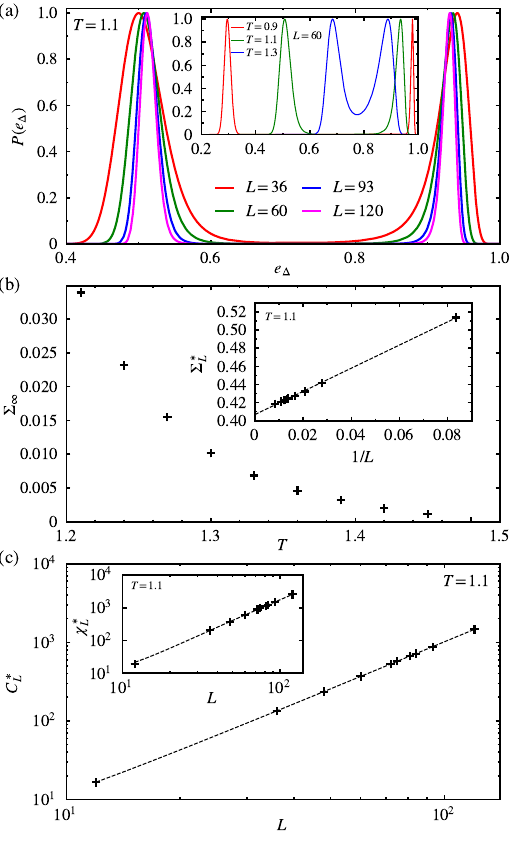}
\caption{(a) Multicanonical probability density functions $P(e_\Delta)$ for a wide range of system sizes at $T=1.1$ (main panel) and for selected temperatures at fixed $L=60$ (inset), illustrating phase coexistence and the rapid growth of the free-energy barrier upon lowering the temperature. (b) Interfacial tension $\Sigma_{\infty}$ as a function of temperature $T$, indicating the onset of the first-order regime near $T\approx1.42$ ($\Delta \approx 0.84$). The apparent approach of $\Sigma_{\infty}$ to zero extends into the second-order region due to finite-size effects associated with the persistent double-peaked energy distribution~\cite{vasilopoulos22}. The inset shows a typical extrapolation yielding $\Sigma_{\infty}\approx 0.41$ at $T=1.1$, about an order of magnitude larger than the values in the main panel, indicating a rapid increase of $\Sigma_{\infty}$ as $T \to 0$. (c) Finite-size scaling of the maxima of the specific heat (main panel) and magnetic susceptibility (inset), consistent with the expected $\sim L^{d}$ scaling for a first-order transition in $d=2$.}
\label{fig:first-order}
\end{figure}

To probe now the remaining first-order transition regime of the phase boundary, emerging at $\Delta \gtrsim 0.85$ ($T \lesssim 1.42$) [Fig.~\ref{fig:first-order}(b)], we carry out multicanonical simulations~\cite{berg92} which efficiently sample configurations separated by large free-energy barriers. In this approach, the Boltzmann weight associated with the crystal-field energy $E_{\Delta}$ is replaced by a generalized weight $W(E_\Delta)$ chosen to produce a flat histogram in $E_\Delta$. For the spin-$1$ Baxter-Wu model, whose density of states depends on two energies $\Gamma(E_J,E_\Delta)$, the multicanonical scheme is applied to $E_\Delta$, allowing unrestricted reweighting in the crystal field $\Delta$ at fixed temperature~\cite{zierenberg15}. The corresponding partition function reads
${\cal Z}_{\rm MUCA}=\sum_{E_J,E_\Delta}\Gamma(E_J,E_\Delta)\,
e^{-\beta E_J}\,W(E_\Delta)$. Once the weights are determined, canonical observables at arbitrary $\Delta$ follow from standard reweighting~\footnote{Multicanonical simulations were performed on an Nvidia Tesla K80 GPU using $26\,624$ workers assigned to independent replicas~\cite{zierenberg13,gross18}.}. This framework enables direct estimates of the normalized probability density functions $P(e_\Delta)$ ($e_\Delta=E_\Delta/L^2$), free-energy barriers, and thermodynamic response functions across the first-order line. In the following we consider system sizes $L \in \{12-120\}$ for temperatures between $T=1.45$ and $T=1.1$. 
In finite systems, a double-peaked structure of $P(e_{\Delta})$ signals the phase coexistence characteristic of a first-order transition~\cite{binder84,binder87}. Representative distributions at $T=1.1$ are shown in Fig.~\ref{fig:first-order}(a) for a wide range of system sizes and display a pronounced suppression of intermediate states. The inset presents $P(e_\Delta)$ at selected temperatures for fixed $L=60$, illustrating the rapid growth of the free-energy barrier upon lowering the temperature. These features provide clear and direct evidence of first-order behavior. From equal-height (eqh) distributions, we extract the free-energy barrier separating the coexisting phases~\cite{lee90,lee91},
$\Delta F_{L}=\frac{1}{2\beta\Delta}\ln\!\left[\left(P_{\rm max}/P_{\rm min}\right)_{\rm eqh}\right]$,
where $P_{\rm max}$ and $P_{\rm min}$ denote the peak value and the intervening minimum of $P(e_\Delta)$, respectively. The corresponding interfacial tension, $\Sigma_{L}=\Delta F_{L}/L$, is expected in two dimensions to scale as $\Sigma_{L}=\Sigma_\infty + c_1 L^{-1} + \mathcal{O}(L^{-2})$ (see the inset of Fig.~\ref{fig:first-order}(b) for a common extrapolation to a nonzero value $\Sigma_\infty=0.407(1)$ at $T = 1.1$). Overall, the interfacial tension increases monotonically upon lowering the temperature, as shown in the main panel of Fig.~\ref{fig:first-order}(b), consistent with earlier studies of the analogous triangular Blume-Capel ferromagnet~\cite{mataragkas25a,mataragkas25b}, although the increase is much more pronounced in the present Baxter-Wu case. Additional confirmation is provided by the scaling of thermodynamic response functions. At a first-order transition, the maxima of the specific heat and susceptibility are expected to scale with the volume, with corrections in inverse powers of the volume~\cite{binder84,challa86,janke97,zierenberg15,zierenberg17}.Accordingly, we fit the full set of data using
$C_L^{\ast} = b_C L^{x}\!\left(1+b_C' L^{-2}\right)$, and
$\chi_L^{\ast} = b_\chi L^{x}\!\left(1+b_\chi' L^{-2}\right)$,
where $\{b_{C}, b^\prime_C, b_{\chi}, b^\prime_\chi\}$ are non-universal amplitudes, as above. As shown in Fig.~\ref{fig:first-order}(c), this Ansatz provides excellent fits for both observables, providing $x = 2.000(5)$ for the specific heat and $x = 2.004(4)$ for the susceptibility, in perfect agreement with the expected value $d=2$. 

\section{Conclusions}
\label{sec:conclusions}

In summary, our combined transfer-matrix and Monte Carlo analysis provides a coherent picture of the phase behavior of the dilute spin-$1$ Baxter-Wu model in the presence of a crystal field. Our results establish the existence of a robust first-order transition regime at low temperatures, characterized by phase coexistence, finite interfacial tension, and the expected finite-size scaling of thermodynamic response functions. Along the continuous transition line emerging from the spin-$1/2$ limit, the central charge remains close to $c\simeq1$, while the effective scaling dimensions deviate systematically as the crystal field increases. Approaching the boundary with the first-order regime, increasingly strong finite-size effects become apparent, rendering the extraction of the asymptotic critical behavior progressively more challenging. Within the resolution of the present study, our results provide no evidence for an isolated multicritical point. Instead, the transfer-matrix spectra, Monte Carlo finite-size scaling, and the progressive loss of conformal scaling are all consistent with an extended crossover region separating the continuous and first-order regimes. The observed evolution of the effective critical behavior is compatible with either genuinely continuously varying critical exponents or a very slow renormalization-group crossover between competing fixed points. Although the present data do not allow us to distinguish conclusively between these scenarios, both naturally account for the broad crossover region observed in the transfer-matrix and Monte Carlo analyses. Distinguishing between these scenarios will require substantially larger system sizes together with further analytical developments beyond the scope of the present work. More broadly, our findings highlight the rich critical behavior generated by the interplay of three-spin interactions, dilution, and the highly degenerate ordered phase, providing a unified framework that reconciles several previously conflicting results for the dilute Baxter-Wu model. From this perspective, the observed crossover appears to be driven not simply by dilution itself but by its nontrivial coupling to the underlying three-spin interaction, providing a possible microscopic explanation for the qualitative differences between the dilute Baxter-Wu and Blume-Capel models.

\begin{acknowledgments}
We are grateful to Professor Per Arne Rikvold for the
very useful correspondence regarding our work. N.G.~F. would like to thank Professor Joao Antonio Plascak for his ongoing collaboration on the problem and the many fruitful discussions over the last years. Part of the numerical calculations reported in this paper were performed at the High-Performance Computing cluster CERES of the University of Essex. The work of A.~V. and N.G.~F. was supported by the  Engineering and Physical Sciences Research Council (grant EP/X026116/1 is acknowledged). D.-H.~K. acknowledges the support from the National Research Foundation of Korea (Grant No. RS-2024-00392445) funded by the Korea government (MSIT).
\end{acknowledgments}

\bibliography{biblio.bib}

\end{document}